# Electroluminescence in *n*-type GaAs unipolar nanoLEDs

**BEJOYS JACOB, JOÃO AZEVEDO, JANA B. NIEDER, BRUNO ROMEIRA**[*]

*INL– International Iberian Nanotechnology Laboratory, Av. Mestre José Veiga s/n, 4715-330, Braga, Portugal*
*[*]bruno.romeira@inl.int*

**In this Letter, we report the observation of electroluminescence (EL) at ~866 nm from *n-i-n* unipolar (electron-transporting) III-V GaAs nanoLEDs. The devices consist of nanopillars with top diameter of 166 nm, arranged in a 10×10 pillar array. Hole generation through impact ionization and Zener tunnelling is achieved by incorporating an AlAs/GaAs/AlAs double-barrier quantum well within the epilayer structure of the *n-i-n* diode. Time-resolved EL measurements reveal decay lifetimes >300 ps, allowing us to estimate an internal quantum efficiency (IQE) higher than 2% at sub-mA current injection. These results demonstrate the potential for a new class of *n*-type nanoscale light-emitting devices.**

**Introduction.** III-V nanophotonic emitting devices, e.g., nanolasers [1,2] and nanoLEDs [3–5], rely on the *p-i-n* architecture. However, as device dimensions shrink to the nanoscale, the resistive *p*-type doping introduces significant drawbacks, including higher resistance [6], optical absorption losses [7], and increased fabrication complexity and costs. A promising approach for nanolight and quantum light sources is the use of *n*-type unipolar (electron-transporting) semiconductors [8]. The use of *n*-doping implies higher carrier mobility, enabling higher speed devices, simplifying the heterostructure band engineering typically needed in *p-i-n* devices, which can reduce fabrication costs.

While the invention of the quantum cascade laser revolutionized mid-infrared unipolar devices [9], achieving room-temperature operation remains a challenge. Notably, there have been limited reports of electroluminescence (EL) in unipolar III-V devices in the near-infrared (NIR) spectrum. Previous studies explored EL in *n*-type double-barrier quantum well (DBQW) resonant tunneling diodes (RTDs), such as GaN/AlN [10], enabled by mechanisms like impact ionization or interband tunneling, provided by the large electric field across the DBQW under applied bias. Although there has been growing interest in unipolar light-emitting devices based on GaN materials, NIR EL in unipolar III-V DBQW structures has predominantly been observed only at cryogenic conditions [11–13], or using *p-i-n* configurations [14]. Room-temperature EL from III-V unipolar devices has so far only been demonstrated in micro-scale devices [15,16], and remains unexplored for sub-µm nanoscale devices.

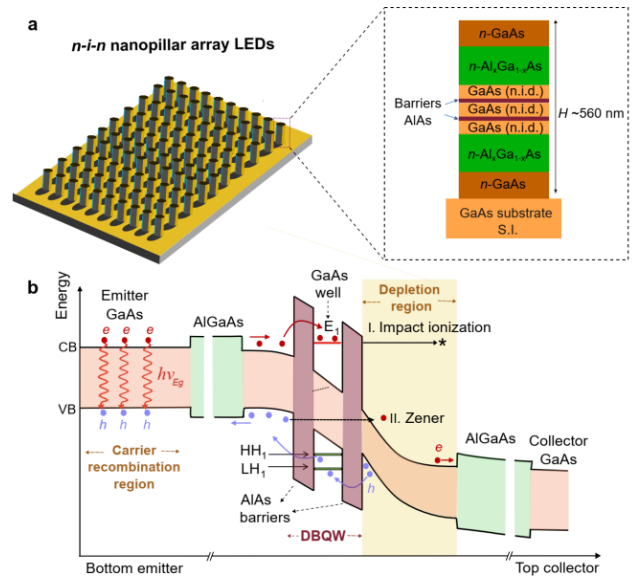

Fig. 1. (a) Schematic of 10×10 nanopillar array LED. The inset shows the *n-i-n* III-V epilayer stack design considered for this study, which includes an AlAs/GaAs/AlAs double-barrier quantum well (DBQW) region. (b) The energy conduction band (CB) and the valence band (VB) diagram under applied voltage, showing: electron (*e*) and hole (*h*) carrier transport, hole generation via I. impact ionization and II. Zener tunneling, the depletion region (highlighted in yellow), and the carrier recombination region on the emitter side producing emission close to the band-edge transition of the GaAs material, $h\nu_{Eg}$~866 nm.

In this Letter, we report room-temperature radiative spontaneous emission at ~866 nm from unipolar *n-i-n* nanoLEDs investigated in a nanopillar array configuration, Fig. 1(a). Hole generation is achieved through impact ionization [17] and Zener (interband) tunnelling [10] by incorporating a 12 nm thick AlAs/GaAs/AlAs DBQW within the epilayer structure of an *n-i-n* diode, as shown in Fig. 1(b). Under applied bias at low voltage, moderate electric fields are achieved across the depletion region where holes are created in the collector region via impact ionization, as highlighted in yellow in Fig. 1(b). These holes accumulating in the collector barrier can go to the emitter region through resonant tunneling or other non-resonant pathways, becoming available for radiative recombination with electrons. At high electric field, enabled by the nanometric DBQW depletion region, additional holes are generated in the GaAs emitter spacer via

Zenner tunneling, Fig. 1(b). Overall, sufficient minority holes become available within the DBQW section for radiative recombination with electrons in the GaAs emitter-doped region, producing emission close to the band-edge transition of the GaAs material, $h\nu_{Eg}$ in Fig. 1(b). We have investigated the EL carrier dynamic properties of electrically modulated unipolar nanoLEDs using time-resolved EL (TREL). Despite the nanometric size of the nanopillar LEDs the results reveal differential carrier lifetimes of >300 ps, which is only 2.8-fold shorter than unipolar microLEDs with identical design. Under low-injection conditions (sub-mA) we estimate an internal quantum efficiency (IQE) ~2% for unipolar nanoLEDs, which can reach up to 6% under high-injection conditions.

**Epilayer design.** The III-V epilayer semiconductor layer stack, Fig. 1(a), of the $n$-type GaAs unipolar nanoLED was grown by molecular beam epitaxy (MBE) on a semi-insulating (SI) GaAs substrate. The epilayer design is identical to the work reported in [15] and consists of an unintentionally doped AlAs/GaAs/AlAs (3.1 nm/6 nm/3.1 nm) layer stack forming the DBQW region. The DBQW was surrounded by thin GaAs spacer layers and $n$-doped-graded AlGaAs cladding layers. Lastly, $n^+$-GaAs ($N=2\times10^{18}$ cm$^{-3}$) layers formed the emitter and collector regions. A thin layer (40 nm) of doped GaAs was chosen for the collector top region to reduce the optical losses caused by light absorption.

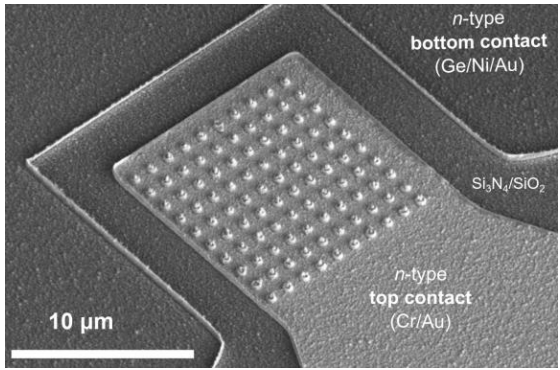

Fig. 2. Scanning electron microscope (SEM) image of a fabricated unipolar nanoLED. The device consists of nanopillars with top diameter of 166 nm, arranged in a 10×10 array.

**Fabrication.** The nanoLEDs were fabricated by defining 10×10 arrays with 1.2 µm pitch. Each pillar had a nominal diameter value of $d_{top}$ ~166 nm, measured from the top pillar. The pillars were fabricated by exposing the nanopillar arrays mask onto a 500 nm thick negative e-beam resist (ARN7520.18) using electron beam lithography. The nanopillar array patterns were then transferred to a 200 nm thick SiO$_2$ layer, which had been initially deposited on top of the cleaned wafer. The nanopillars were deep etched down to approximately $H$ ~560 nm via reactive ion etching (RIE) using chlorine chemistry. It should be noted that the RIE recipe used subtend an angle of $\theta_c$~16⁰ to the normal of the bottom surface forming a cone shaped nanopillar. Following etching, the nanopillars were coated with a ~10 nm thin layer of Si$_3$N$_4$ deposited by low-frequency plasma after a sulphur chemical treatment, using the same procedure as described in [18], thus enabling both electrical isolation and surface passivation of the nanopillars. An additional 100 nm layer of SiO$_2$ was deposited by high-frequency plasma vapor deposition. Via openings on top of the nanopillars coated with the Si$_3$N$_4$/SiO$_2$ dielectric layer were formed using a planarized etch-back process. The $n$-type bottom and top contacts were formed using two optical lithography steps, followed by metal sputtering with Ge/Ni/Au and Cr/Au metal alloys, respectively. To enable light extraction from the nanopillar sidewalls, the top metal contacts were sputtered onto the sample at a 45° angle. A scanning electron microscope (SEM) image of a fabricated nanoLED array is shown in Fig. 2.

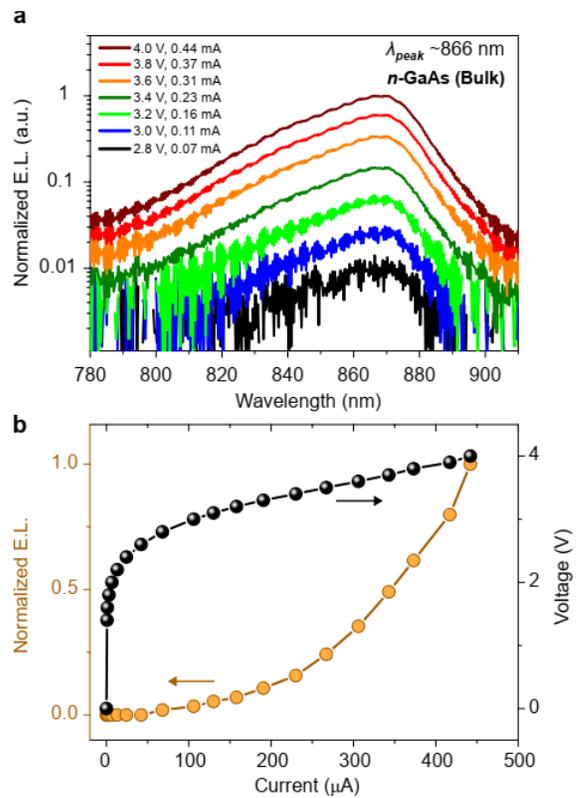

Fig. 3. Experimental room-temperature static characteristics of the unipolar nanoLED. (a) Electroluminescence spectra as a function of the electrical pumping conditions. (b) Light-current-voltage (*L-I-V*) characteristics.

**Electroluminescence.** The experimental static characteristics were measured at room temperature using a ground-signal-ground coplanar waveguide electrical probe (10 GHz bandwidth) to electrically connect the unipolar nanoLED devices. The DC bias was provided with a source meter (Keithley, model 2280S) connected to a high-bandwidth bias-T. The EL from the nanopillar LED arrays was collected by a multimode lensed fiber (spot size ~25 µm) connected to a fiber spectrometer with a slit size of 25 µm (Avantes, AvaSpecHSC1024x58TEC-EVO-new). Fig. 3(a) shows a typical EL spectra collected for 20 s as a function of the bias current for an applied voltage ranging from 2.8 to 4 V. The unipolar nanoLED displays a prominent EL emission signature at 866 nm. NanoLEDs of other sizes show identical EL peaks (Supplement 1, Fig. S1). The fringe-like features in the spectra are related to second-order effects from the spectrometer grating, which can be mitigated by using an order-sorting filter. The broad emission peak, peaking at 866 nm, corresponds to the band-edge transition at room-temperature from the GaAs-doped layers on the bottom emitter side. Notably, in previous work on unipolar microLEDs [15], emission ~806 nm from the quantum well within the DBQW active center region (Supplement 1, Fig. S2) was reported for micropillars with diameters above 5 µm. However, this emission is absent in the EL spectra from nanoLEDs shown in Fig. 3(a) and from other samples fabricated with nanopillars smaller than 1 µm (Supplement 1, Fig. S1 and S4). We attribute this to the surface recombination effects in the etched GaAs nanopillars, where non-radiative recombination becomes more important as the surface-to-volume ratio of the pillars substantially increases [18]. Considering the small size of the pillars, other effects may contribute to the other pathways for current, such as hot electrons, thermal activation, and sidewall leakage, resulting in the complete absence of the DBQW 806 nm peak.

Fig. 3(b) shows the experimental light-current-voltage ($L$-$I$-$V$) characteristics. The $I$-$V$ exhibits a diode-like turn-on voltage at ~2.8 V (Supplement 1, Fig. S2), which is related with the high series resistance (~2.85 kΩ). This results mainly from the small contact area of the nanopillar devices, but also contribution from the AlGaAs layers and the non-optimized metal contacts. Considering the small contact area of the devices, the nanoLEDs operate under a large current density ranging from 1 kA cm$^{-2}$ to 10 kA cm$^{-2}$. The $L$-$I$ curve indicates a non-linear, gradual increase in electroluminescence as the current increases, suggesting changes in the device efficiency as a function of electrical injection. We attribute this to a combination of i) non-radiative surface effects at lower current injection conditions which reduces EL, and ii) additional hole generation mechanism via Zener interband tunnelling for higher electric field, which improves hole injection efficiency at higher voltage conditions.

**Time-resolved electroluminescence.** Time-resolved EL (TREL) measurements were used to estimate the differential carrier lifetimes of the EL of the unipolar microLEDs. The unipolar microLEDs were electrically modulated by voltage signals using a pulse generator connected to the radio frequency (RF) port of a high-bandwidth bias-T, and DC biased through a source meter. The electrical pulses consisted of square wave signals with a repetition rate of 1 µs, a pulse width of 10 ns, and rise and fall times ~70 ps, with a peak-to-peak voltage of 1 V (Supplement 1). The rise and decay curves obtained for the unipolar nanoLED of Fig. 2, as a function of the DC voltage, are shown in Fig. 4 and the decay curves were fitted as described in Supplement 1. As shown in Fig. 4, the carrier differential lifetime for low injection conditions is ~300 ps. The sub-ns value indicates that the non-radiative recombination are dominant, specifically surface recombination, due to the high surface-to-volume ratio of the nanopillars fabricated. However, we notice the 300 ps value is only 2.8-fold shorter than the lifetimes measured in unipolar microLEDs under low injection conditions (~821 ps for a 6 µm wide micropillar, Supplement 1, Fig. S3), despite the size of the nanopillars is 24-fold smaller. As discussed next, we attribute this long lifetime to an improved surface passivation effect related to the sulfur chemical treatment followed by the 10 nm $Si_3N_4$ coating layer deposited by low-frequency plasma deposition employed in our nanopillars, whereas microLEDs used $SiO_2$ coating deposited by high-frequency plasma deposition shown to provide a poor passivation effect [18].

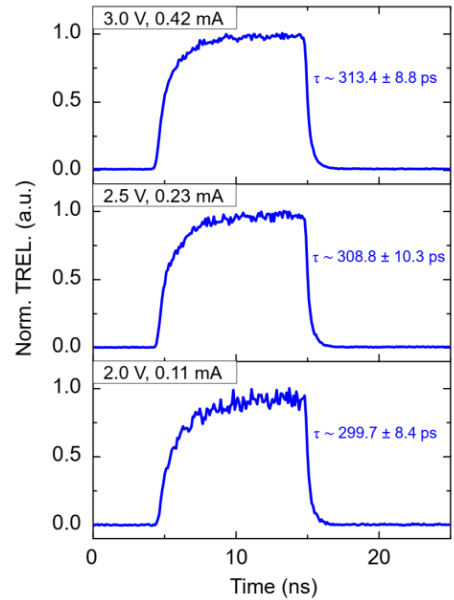

Fig. 4. Experimental time-resolved electroluminescence (TREL). The unipolar nanoLED is modulated with electrical pulses with a pulse width of 10 ns and a peak-to-peak voltage of $V_{pp}$=1 V. It is shown the differential carrier lifetime (τ) as a function of the DC bias: (a) 2 V, (b) 2.5 V, and (c) 3 V.

**Surface recombination velocity and internal quantum efficiency (IQE).** To evaluate the IQE (Supplement 1) of the unipolar nanoLEDs, we first quantify the surface velocity, $S$. We assume that under the low excitation conditions employed in the experiments of Fig 4, the surface-related non-radiative recombination rate scales as $4S/d_a$, so that $S$ can be estimated directly from the size-dependent (diameter of the active region, $d_a$) carrier lifetime in the low injection regime. Applying $\tau^{-1} \approx 4S/d_a$ to the results in Fig. 4, the calculated surface velocity recombination for the nanoLED ranges from $2.08 \times 10^4$ to $1.99 \times 10^4$ cm s$^{-1}$. These values are comparable to the results achieved for optically-pumped GaAs nanopillars using an identical surface passivation method ($S=1.1 \times 10^4$ cm s$^{-1}$ [18]), indicating an improved passivation also for electrically pumped nanopillars.

To calculate IQE we use Eq. (E.1) (Supplement 1) and assume the parameter $A=3.34 \times 10^9$ s$^{-1}$ (taking $S=2.08 \times 10^4$), and parameters $B=1.8\times10^{-10}$ cm$^3$ s$^{-1}$, and $C=3.5\times10^{-30}$ cm$^6$ s$^{-1}$ taken from GaAs material. For simplicity of analysis, we assume the charge densities are in the range of $N_d =2\times10^{17} – 2\times10^{18}$ cm$^{-3}$, corresponding to the $n$-type doping concentration in the GaAs emitter side. Fig. 5a) shows the calculated IQE values for the unipolar nanoLED with $d_{top}$~166 nm and a comparison with the case of a 6 μm wide unipolar microLED (considering a higher surface recombination velocity of $S$ ~$2\times10^5$ cm s$^{-1}$ [15]). It should be noted that since the radiative emission is generated in the bottom GaAs layers, for the calculation we have used the estimated size of the pillar bottom emitter taking into account the cone-shaped pillar ($d_a = d_{top} + 2 H \tan\theta_c$ =490 nm). In Fig. 5, we show the results assuming a low injection efficiency value of $\eta_I=0.11$, typical of GaAs-based unipolar micro-sized devices reported so far [15]. We estimate an IQE of 2% for our nanoLEDs under low pumping conditions ($N_d = 2\times10^{17}$–$2\times10^{18}$ cm$^{-3}$). Considering higher pumping conditions ($N_d=2\times10^{19}$ cm$^{-3}$) and just before saturation due to Auger recombination, the IQE can reach up to 6%. The quantum efficiency can be further increased by using optimized devices with higher carrier injection efficiency ($\eta_I>0.2$ as suggested in [16]) leading to an IQE increase to IQE>0.15, Fig. 5b). Lastly, the estimated EQE of our unipolar nanoLEDs (<10$^{-5}$, Supplement 1) is in line with previous $p$-$i$-$n$ III-V nanoscale light emitting devices which report EQEs ranging between 10$^{-6}$ and 10$^{-4}$, Supplement 1.

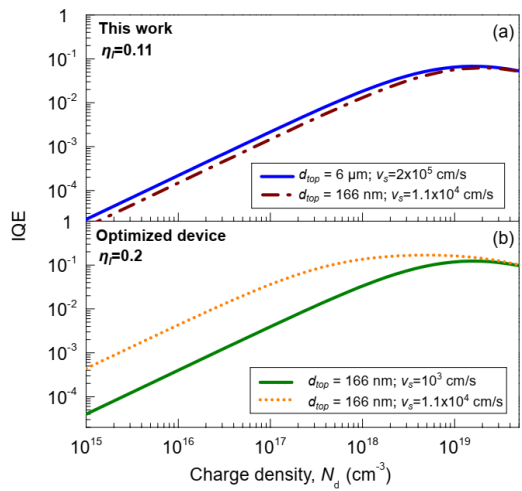

Fig. 5. (a) Calculated internal quantum efficiency (IQE) for the characterized unipolar nanoLED (blue continuous line) and the example of an unipolar microLED [15] (brown circles continuous line). (b) Calculated IQE assuming optimized unipolar nanoLED with improved injection efficiency $\eta_I= 0.2$ [16], and further improved surface recombination ($S$ ~10$^3$ cm s$^{-1}$) [19]

**Conclusion.** We have demonstrated room-temperature electroluminescence at ~866 nm from $n$-type GaAs unipolar nanoLEDs with nanopillar top diameters as small as 166 nm, achieved via hole generation through impact ionization and Zener effects. This novel light-emitting nanodevice exhibits differential carrier lifetimes only 2.8-fold shorter than the lifetimes measured in unipolar micro-sized LEDs, indicating a reasonably high quantum efficiency. This $n$-type source can provide an approach to mitigate electrical contact resistance and optical absorption losses resulting from $p$-doped materials. This substantially simplifies the heterostructure typically required in $p$-$i$-$n$ devices. The results could be extended to other III-V compound semiconductors (e.g., InP) covering visible and infrared wavelengths for a new class of $n$-type optoelectronic and quantum light source [8] devices.

**Funding.** European Union, H2020-FET-OPEN framework programme, Project 828841 – ChipAI, Horizon Europe, project 101046790 – InsectNeuroNano, and Fundação para a Ciência e a Tecnologia (FCT) project 2022.03392.PTDC – META-LED.

**Acknowledgment.** We acknowledge access to the Micro and Nanofabrication Facility and the Nanophotonics and Bioimaging Facility at INL. We acknowledge Jérôme Borme, INL, for the e-beam lithography, and José Figueiredo, Universidade de Lisboa, on the device design discussion. BJ acknowledges the support in frame of the PhD program in Electrical Engineering, Electronics and Automation at Carlos III University of Madrid, Spain.

**Disclosures.** The authors declare no conflicts of interest.

**Data availability.** Data underlying the results presented in this paper are not publicly available at this time but may be obtained from the authors upon reasonable request.

**Supplemental Document.** "See Supplement 1 for supporting content.

**Supplement 1.**

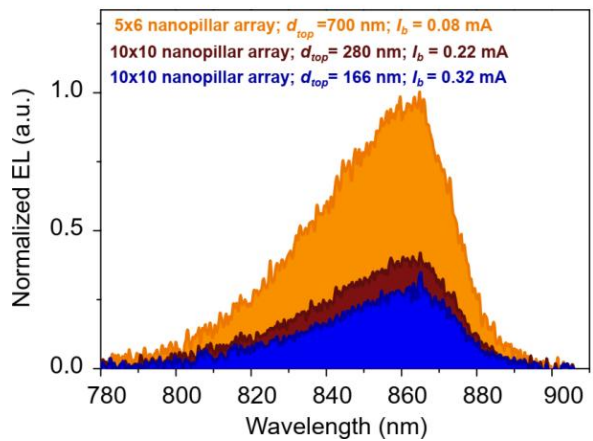

Fig. S1. Electroluminescence (EL) spectra comparison between unipolar nanoLED arrays of top diameter $d_{top}$ = 700 nm (5×6 array), $d_{top}$ = 280 nm (10×10 array), and $d_{top}$ = 166 nm (10×10 array). $I_b$ corresponds to the bias current conditions.

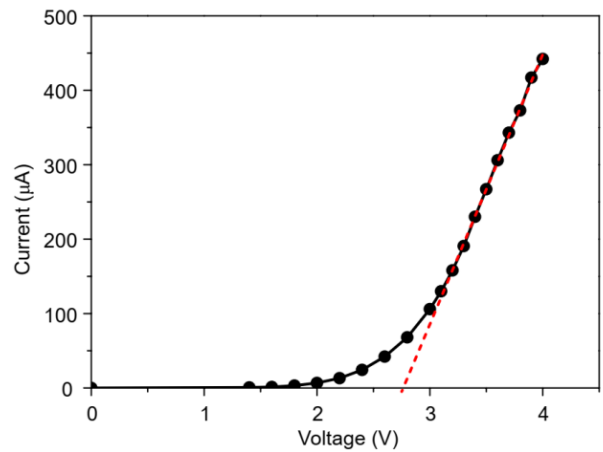

Fig. S2. I-V characteristic of a unipolar nanoLED array (10×10 array) of top diameter $d_{top}$ = 166 nm. The linear fit indicates a turn-on voltage ~2.8 V which agrees with the measured EL spectra in Fig. 3 of the main manuscript.

**Electroluminescence in unipolar micro- and nanopillars.**
Figure S3 shows the peak emission signatures for unipolar microLEDs for larger micropillar diameters (6 µm, 8 µm, and 10 µm). A detailed description of the various spectral signatures is provided in our previous work [1] using the same epilayer design for microscale devices. Figure S4 shows the trend of the emission of unipolar pillar LEDs with diameters ranging from 10 µm down to 166 nm. As shown in Fig. S4, for larger sizes (from 6-10 µm), the DBQW 806 nm emitting starts to be less pronounced for a size of 6 um. For sizes of 1 µm or below (Fig. S1 and S4), the DBQW 806 nm peak is completely absent. We note the peak emission wavelengths for the devices analyzed here did not change with the pumping conditions. The peak emission from the DBQW is more pronounced for larger micropillar diameters. This behavior is attributed to the larger device size, which enhances impact ionization and directs holes into resonant pathways defined by the DBQW, enabling efficient radiative recombination in the QW region. As the pillar size decreases, the contribution from the DBQW diminishes, and more recombination occurs in the *n*-GaAs bottom contact due to the reduced availability of resonant pathways along the pillar.

We note to achieve emission from the QW, coherent current, that is, carriers that are injected in the QW via resonant tunneling effect, should dominate over other pathways for carrier transport. However, considering the small size of the pillars, other effects may contribute to the other pathways for current, such as hot electrons, thermal activation, and sidewall leakage [2], resulting in the complete absence of the DBQW 806 nm peak.

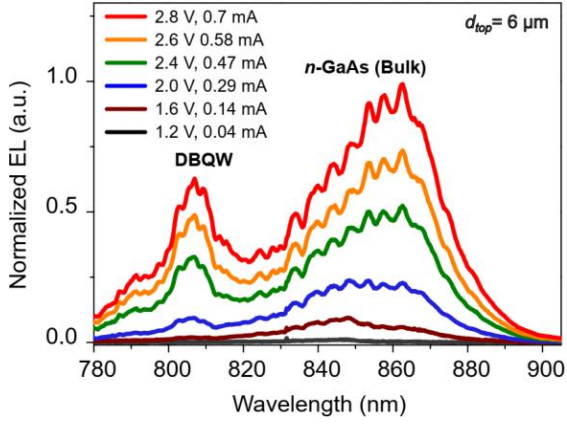

Fig. S3. Unipolar microLED device (top diameter $d_{top}$ =6 μm) under different bias conditions. Unlike the unipolar nanoLED, the microdevice exhibits EL signatures from both the double barrier quantum well (DBQW) emitter ($hv_{DBQW}$ ~806 nm) and the $n$-GaAs bulk emitter at $hv_{Eg}$ ~866 nm [1].

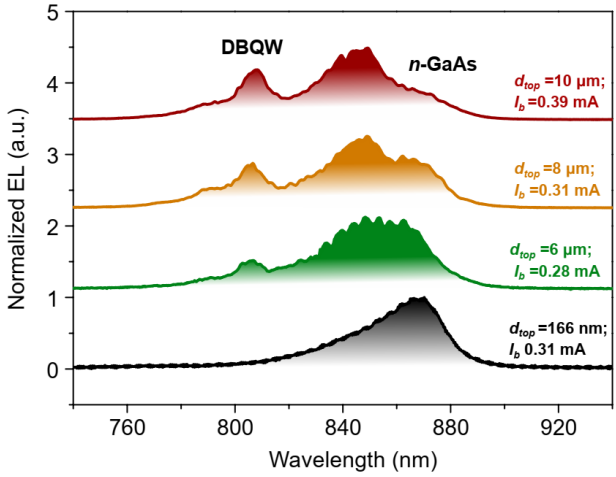

Fig. S4. Comparison of micro scale LED device (top diameter $d_{top}$=6 μm, 8 μm, and 10 μm,) with under $d_{top}$= 166 nm nanopillar array LED with respective similar puping conditions. Unlike the unipolar nanoLED, the microdevice exhibits EL signatures from both the double barrier quantum well (DBQW) emitter ($hv_{DBQW}$ ~806 nm) and the $n$-GaAs bulk emitter at $hv_{Eg}$ ~866 nm [1].

**Time-resolved EL (TREL) setup.** The modulated optical output signal from nanoLED was collected by a lensed fiber (as described for the static characterization), filtered by a bandpass filter with 40 nm full width at half-maximum (FWHM) centered at 850 nm, and connected to a fiber-coupled single-photon counting avalanche photodetector (SPAD) (MPD, model PDM series) with a temporal resolution of ~50 ps. The SPAD was connected to a correlation card (Becker & Hickl, TCSPC 150N) controller which measured the time between the arrival of a luminescence photon at the SPAD (the start signal) and the excitation signal arriving from the trigger signal of the pulse generator (stop signal). The photon arrival times are then binned to obtain a histogram that corresponds to the time-dependent output intensity of the electrically modulated nanoLEDs. The differential carrier lifetimes shown in Fig. 4 were estimated employing a mono-exponential fit of the measured decay curves, based on the following function:

$$N_{ph} = N_a e^{-\left(\frac{t-t_0}{\tau}\right)} + N_0 \quad (E.1)$$

where $N_{ph}$ represents the normalized number of photons, $N_a$ is the amplitude of the exponential fit, $N_0$ is the offset value, and $\tau$ represents differential carrier lifetime.

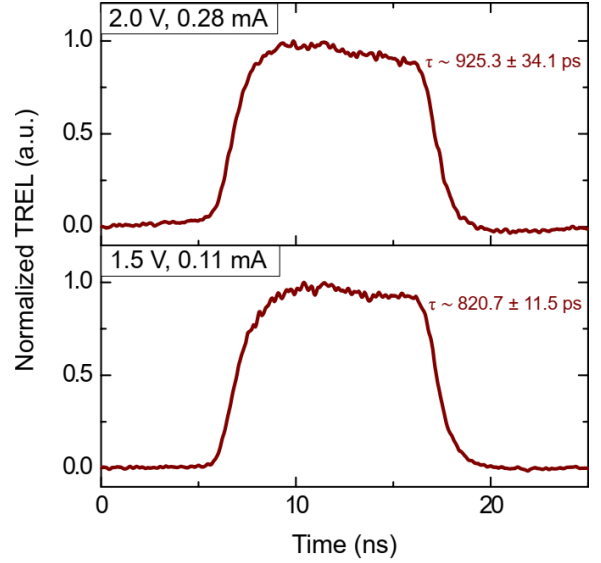

Fig. S5. Experimental time-resolved electroluminescence (TREL) of the unipolar $d_{top}$ =6 μm microLED reported in [1] electrically modulated with electrical pulses with a pulse width of 10 ns and a peak-to-peak voltage of $V_{pp}$=1 V. The results show the differential carrier lifetime ($\tau$) as a function of the DC bias: (a) 1.5 V, and (b) 2.0 V

**Internal quantum efficiency (IQE).** The IQE is given by:

$$\text{IQE} = \eta_r \eta_i \quad (E.1)$$

where $\eta_i$ represents carrier injection efficiency, and $\eta_r$ is the radiative recombination efficiency, which is given by:

$$\boldsymbol{\eta_r} = \frac{BN_d}{(A + BN_d + CN_d^2)} \quad (E.2)$$

where $B$ is the bimolecular recombination coefficient, $C$ is the Auger recombination coefficient, and $N_d$ is the local carrier density of donors. The parameter $A$ is given by: $\boldsymbol{A} = \frac{4S}{d_a}$, where $S$ is the surface recombination velocity, and $d_a$ is the diameter of the active emitting region of the nanopillar.

**External quantum efficiency (EQE).** The EQE is given by:

$$\text{EQE} = \eta_c \text{ IQE} \quad (E.3)$$

$\eta_c$ is the coupling coefficient and is determined by the numerical aperture (NA~0.2) of the lensed fiber used to collects the EL data, ratio of extraction efficiency of the pillar ($\eta_{pillar}$) compared to the planar LED emission ($\eta_{bulk}$) and ratio of light coupled with metal coated and uncoated pillars ($\gamma$).

$$\eta_c = \frac{1}{4}\left(\frac{NA}{n}\right)^2 \left(\frac{\eta_{pillar}}{\eta_{bulk}}\right)\gamma \tag{E.4}$$

where $n \approx 3.55$ refractive index of GaAs.

Following the similar methodology as in previous work [2], $\eta_{pillar}$ was determined. By assuming the light extraction efficiency of the largest pillar diameter $d = 5$ µm is same as $\eta_{bulk} \approx 0.02$, and using the photoluminescence (PL) intensity per effective area plot for GaAs/AlAs nanopillars (see figure S6), we can estimate the $\eta_{pillar}$ for smaller pillars. The PL experiments were performed in a confocal system (WITec Alpha 300R), using a 532 nm laser source operated at 150 µW, and the emission from the pillars was collected through an objective lens with NA=0.9. The collection path includes a UHTS300 spectrometer with a 600 lines/mm diffractive grating coupled to a Peltier cooled CCD detector. Since all pillars have a tapered shape, with a tilted angle $\theta_c \sim 16°$ relative to the vertical direction, their lateral emitting area is regarded as non-negligible for submicron cases ($d \leq 500$ nm). This allows for a realistic and reliable estimate of the actual PL intensity per area.

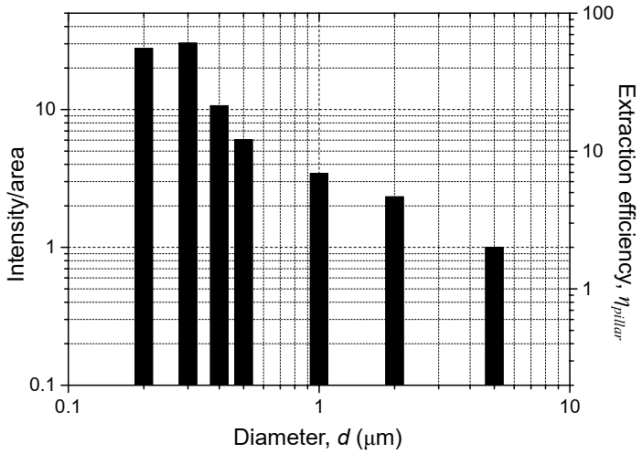

Fig. S6. Integrated intensity per effective emitting area as a function of diameter for micro- and nanoscale pillars ($d$ = 200 nm to $d$ = 5 µm).

To determine γ ratio of the integrated PL measurements (see Fig. S7), we considered the PL integrated intensity with and without metal coating for $d_{top}$=240 nm pillar and γ~1/47.

Based on the above analysis of PL size-dependent measurements (Fig. S6), the light outcoupling efficiency of uncoated pillars is estimated to be ~56%. This agrees with our previous work showing extraction efficiency from undoped GaAs/AlGaAs nanopillars in the range of 33%-57% [2]. This large extraction (as compared to the low extraction efficiency in planar microLEDs ~2%), is mainly related to the strong reduction of the total internal reflection effect for nanopillars with sizes below 1 µm. However, in this work the electrically pumped nanoLEDs are coated with metal layers for the electrical contacts. Although a ~45° angle deposition of the metals was made to leave one of the sides of the nanopillars uncoated, PL measurements comparing uncoated and coated nanopillars reveal that the emission drops by a factor of 47 (Fig. S7), in which a fraction of this PL drop is related to the reduction of the illuminated area of the metal-coated nanopillar. Considering the worst-case scenario that the emission of our coated nanopillars is reduced 47-fold, the outcoupling efficiency for electrically connected nanoLEDs is of the same order of magnitude as the IQE. As a result, considering also the limited NA aperture of the lensed fiber to couple light for the nanopillars the estimated EQE (Equation E.4) for our devices is much lower than the reported IQE (EQE<$10^{-5}$). We note this estimated EQE is comparable with previous reported p-i-n III-V nanoscale light emitting devices, including p-i-n nanowire LED [3], p-i-n nanopillar metal-dielectric LED [4], p-i-n quantum dot PhC LED [5], which report EQEs ranging between $10^{-6}$ and $10^{-4}$.

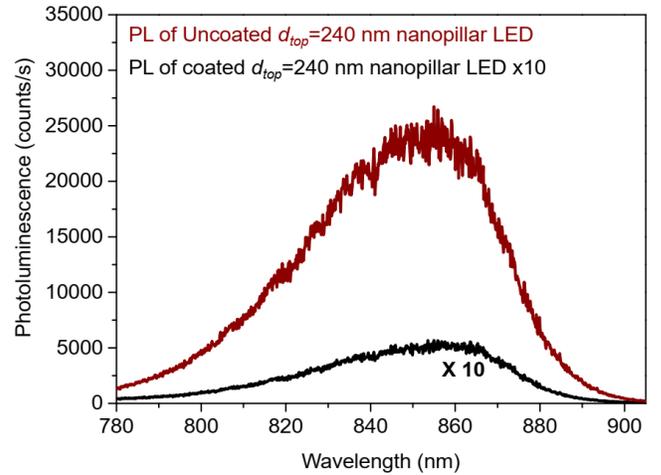

Fig. S7. PL measurements of nanopillar LED with $d_{top}$ = 240 nm with and without top metal coating. The PL measurement with of uncoated nanopillar LED is multiplied by 10 for better visualization.

We conclude the main limitation for the EQE in our devices, besides potential improvements in injection carrier efficiency and further surface recombination improvements, is the low extraction efficiency due to the metal coating which can be improved by either optimizing the angle deposition method (as for example increasing the pillar height, reducing the tapering effect, or decreasing the pitch of the nanopillars as reported in [6]) or using ITO transparent contacts.

**References (Supplement 1)**